# Are insight problems really different from noninsight problems?

Research proposal CKI-B

Matthijs Melissen
Student Number: 0423165

**Abstract**
*In this text, I will suggest an electroencephalogram (EEG) experiment with which it will be possible to see whether there is biological evidence for the frequently made distinction between insight and noninsight problems. What is meant with insight here is the 'aha'-experience, the sudden discovery of how a problem works.*
*First, I will give a summary of the research done by Auke Pols in his thesis 'Insight in problem solving' (Pols, 2002), an introductory text on insight. This part of this text consists of an overview of the questions Pols asks himself, the answers to these questions, and the methods he uses to find them. Secondly, I will formulate my own research question, and propose the methods with which I want to answer this question.*

Pols set himself multiple goals. First, he reviews the history of insight research. Secondly, he wants to give a definition of the words 'insight' and 'insight problems'. Thirdly, Pols tries to find a model which explains insight. Fourthly, the opinions of philosophers about insight are being reviewed. Fifthly, Pols investigates the relation between Artificial Intelligence and the concept of insight.

There are two important movements in the area of insight research: on the one hand the associationists, who claim that the mind consists of ideas (elements) linked by associations (links). According to this model, thinking is in effect moving from idea to idea via a chain of associations. Insight in problem solving is, according to associationism, just a specific form of problem solving, what is in itself a specific form of stimulus-response association. On the other hand there are the Gestaltists, who theorized that the 'flash of insight' wasn't so much the discovery of the solution to a problem, as a kind of 'restructuring' of the problem that opened the way to the solution.

Pols defines the process of getting 'insight' in a problem as a (1) sudden (2) transition (3) which causes a deeper or more appropriate form of understanding the problem. The best way of defining insight problems is by giving a taxonomy or classification of insight problems. It is possible to divide them into routine- and nonroutine problems, object-use and spatial-insight problems, pure and hybrid problems or into conceptual access problems and novel representation problems. The fact that both the problem and the solution can be open and closed is another way to divide the problems. Unfortunately, most classes also contain noninsight problems.
The best definition is considered to be the one of Schooler, Ohlsson and Brooks (1993): "an insight problem is well within the competence of the average subject, has a high probability of leading to an impasse and has a high probability of 'rewarding' sustained effort with an aha-experience."

One of the models of insight problem solving is the constraint model of Isaak & Just (1995). Constraints are a kind of mental rules that prescribe what possibilities should and should not be considered when searching for a solution. They can be explicitly stated in the problem, or implicit, (often unconsciously) self-imposed strategies of the problem solver: assumptions of what should and should not be done, although it is mentioned nowhere. These implicit constraints are often helpful. Insight problems, however, seem designed specifically to evoke the application of inappropriate implicit constraints. To solve the insight, these constraints have to be overcome: the moment of insight. The constraint model does not go into detail on what happens between the fruitless search for the solution and the realization and removal of implicit constraints, however. To study that phase, we can use the spreading activation theory. This theory assumes that our knowledge is represented in a kind of subconscious semantic network, consisting of concepts linked to each other. Some concepts that are closely related also have stronger links, while concepts that are only weakly related have much weaker links. When a problem is read, the concepts presented in the problem are activated, and activation spreads through the network. The strongest link draws the most attention. Insight problems, however, can only be solved by activating weakly linked concepts. After some time, these weak links will be activated, but it will often take longer until the activity of the strong but incorrect activations has subside enough to allow for the weak activations to enter consciousness. So gaining insight could be a gradual process for the subconscious, but a sudden event for consciousness, since the correct activations may only enter consciousness when the activation exceeds a certain threshold.
Of course, alternative models for insight in problem solving have been proposed. In the original article is dealt with the classical associationist model, the evolution model, the fixation model and the mental ruts model. Pols considers the constraint model still less flawed and yet robust enough to encompass all problems that are denoted insight problems by the chosen definition.

Pols does not say much about the philosopher's opinion of insight. The only philosopher he mentions is Wittgenstein. Wittgenstein proposes a belief network consisting of strong beliefs (axioms) and weak beliefs (gossips) and everything in between. New information always changes a set of beliefs, since new information, apart from being an addition itself, always illuminates or contradicts other believes. This gives Wittgenstein network both associationist and Gestaltist properties: it is a belief network in which concepts are linked, yet when a new belief enters, a whole portion of the network is immediately changed. Unfortunately, Wittgenstein does not go deeper into the properties of such a network or its effect on problem solving.

Neural networks could be used for helping a subject noticing invariants: the aspects of a problem that remain unchanged during different solution attempts, and that often are the crucial features of a problem. Further, they could be used to model mechanisms like spreading activation and fixation. And classification networks could be used to try to classify insight problems or divide a method to reliably distinguish insight problems from noninsight problems.
The use of evolutionary algorithms could also be useful. Both in neural networks ans in evolutionary algorithms, the greatest problem seems to be how to represent a problem and its solution as bit strings. When that problem is solved, however, evolutionary algorithms could be used to model a number of processes related to insight problems, like fixation on wrong solution paths, what happens during the incubation phase and spreading activation.

## *Research proposal*

A lot is said about insight problems, but nowhere is proven that there is really something special about these kind of problems. The question I want to answer is, Are insight problems really different from noninsight problems? This question is also related to the discussion between the associationists and Gestaltists (see page 1).

The use of brain imaging techniques is one of Pols recommendations for further research. It seems a good idea to use a brain imaging technique to answer the question I proposed.

There has already been research to this question by making use of Magnetic Resonance Imaging (MRI) techniques (Beeman and Bowden, 2000), but I think it is better to use electroencephalography (EEG), because the 'flash of insight' happens in a very short time, and the temporal resolution of MRI is too coarse to detect this well. I guess EEG is a more appropriate technique to 'catch' the moment of insight.

Because I don't expect we will measure large effects, it will be necessarily to amplify the results by making use of Event Related Potentials (ERP). It would be ideal to 'lock' the ERP on the moment of insight, the 'aha'-experience. However, this is impossible, because we don't know exactly at what moment the 'aha'-experience happens. Asking the subject to press a button when the insight moment has 'arrived' is difficult, because it is hard to give the subject instructions which will make this clear, without telling then that the experiment deals with insight (which might alter their solution strategy). To overcome this problem, we will ask subjects to press a button when they find a solution, use response-locked ERP, and use a problem in which insight leads immediately to the solution. Another constraint is that the presentation of the problem to the subjects must not immediately make the solution clear, because then we can't make the difference between the activation caused by the presentation of the problem and the activation caused by the finding of the solution. The famous 9-dot problem (Weisberg, 1995) doesn't fulfil the first constraint, so I created a variation on this problem, which is expected to fulfil both constraints.

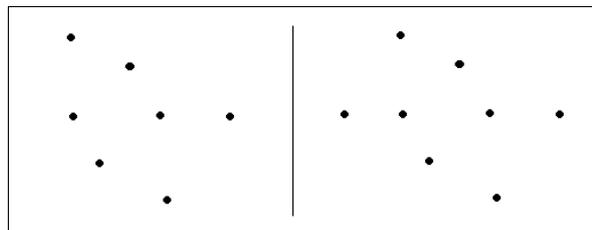

*Image 1: Left: the insight problem used in this experiment. All dots have to be connected by three straight lines without lifting a pen from the paper. Right: the control problem. The same instructions hold.*

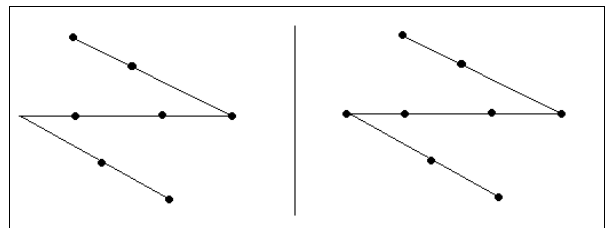

*Image 2: The solutions to the problems shown in image 1.*

The subjects will be divided into two groups. The first group will be presented the left image from image 1, the insight problem. The second group, a control group, will be shown the right image (the control problem). Members of both groups are asked to mentally connect the dots with three straight lines without lifting a pen from their paper, and push a button when they are ready. The solutions to both problems are shown in image 2. Note that the insight problem and the control problem have the same solution.
During this task, an EEG will be recorded. As mentioned before, the individual EEG scan results will be converted into a response-locked ERP.

When activity during 'the moment of insight' is located in another place as activity during the rest of the problem solving, there is evidence that insight problems are really different from other problems. If this happens, the experiment will also give evidence about what part of the brain deals with insight problems and the 'aha'-experiences.

**References**

Beeman, M.J. & Bowden, E.M. (2000). The right hemisphere maintains solution-related activation for yet-to-be-solved problems. *Memory & Cognition, 28(7)*, 1231-1241.

Pols, A. J. K. (2002). Insight in problem solving. Preprint in the CKI-scriptieserie, http://www.phil.uu.nl/preprints/scripties.

Weisberg, R.W. (1995). Prolegomena to theories of insight in problem solving: a taxonomy of problems. In R.J. Sternberg & J.E. Davidson (Eds.), *The nature of insight*, 157-196. MIT Press. Quoted in Pols (2002).